\documentstyle[12pt,epsf,epsfig]{article}
\def\ga{\mathrel{\raise.3ex\hbox{$>$\kern-.75em\lower1ex\hbox{$\sim$}}}}
\def\la{\mathrel{\raise.3ex\hbox{$<$\kern-.75em\lower1ex\hbox{$\sim$}}}}
\def\be{\begin{equation}}
\def\ee{\end{equation}}
\def\ba{\begin{eqnarray}}
\def\ea{\end{eqnarray}}
\def\ga{\mathrel{\raise.3ex\hbox{$>$\kern-.75em\lower1ex\hbox{$\sim$}}}}
\def\la{\mathrel{\raise.3ex\hbox{$<$\kern-.75em\lower1ex\hbox{$\sim$}}}}

\newcommand{\bi}[1]{\bibitem{#1}}
\newcommand{\fr}[2]{\frac{#1}{#2}}

\begin{document}

\baselineskip=16pt
\begin{titlepage}
\rightline{UMN--TH--2103/02}
\rightline{TPI--MINN--02/18}
\rightline{SUSX--TH/02-011}
\rightline{hep-ph/0205269}
\rightline{May 2002}
\begin{center}

\vspace{0.5cm}

\large {\bf
Constraints on the Variations of the Fundamental Couplings}
\vspace*{5mm}
\normalsize

{\bf Keith A. Olive}$^{1,2}$, {\bf Maxim Pospelov}$^{3}$, {\bf
Yong-Zhong Qian}$^{2}$,\\ {\bf Alain Coc}$^4$, {\bf Michel Cass\'e
}$^{5,6}$,  and {\bf Elisabeth Vangioni-Flam}$^5$

\smallskip
\medskip

$^1${\it Theoretical Physics Institute, School of Physics and
Astronomy,\\  University of Minnesota, Minneapolis, MN 55455, USA}

$^2${\it School of Physics and
Astronomy,\\  University of Minnesota, Minneapolis, MN 55455, USA}

$^4${\it CSNSM, IN2P3/CNRS/UPS, B\^at 104, 91405 Orsay, FRANCE}

$^5${\it IAP, CNRS, 98 bis Bd Arago 75014 Paris, FRANCE}

$^6${\it SAp, CEA, Orme des Merisiers, 91191 Gif/Yvette CEDEX, FRANCE}

\smallskip
\end{center}
\vskip0.6in

\centerline{\large\bf Abstract}

We reconsider several current bounds on the variation of the
fine-structure constant in models where all gauge and Yukawa couplings
vary in an interdependent manner, as would be expected in unified theories.
In particular, we re-examine the bounds established by the Oklo
reactor from the resonant neutron capture cross-section of $^{149}$Sm. By
imposing variations in
$\Lambda_{QCD}$ and the quark masses, as dictated by unified theories,
the corresponding bound on the variation of the fine-structure constant
can be improved by about 2 orders of magnitude in such theories.  In
addition, we consider possible bounds on variations due to their effect
on long lived $\alpha$- and
$\beta$-decay isotopes, particularly $^{147}$Sm and $^{187}$Re. 
 We obtain a strong constraint on $\Delta \alpha / \alpha$, comparable to
that of Oklo but extending to a higher redshift corresponding to the age
of the solar system, from the radioactive life-time of
$^{187}$Re derived from meteoritic studies.
We also analyze the astrophysical consequences of perturbing the decay $Q$
values on bound state
$\beta$-decays operating in the $s$-process.

\vspace*{2mm}
%\smallskip\newline

\end{titlepage}

\section{Introduction}

The nature of fundamental constants in physics is a long-standing problem.
While certain constants can be thought of as merely unit conversions ($c,
\hbar$, $k_B$, etc.), others such as gauge and Yukawa couplings can
be thought of as dynamical variables. Indeed, such is the case in
string theory, where the only fundamental parameter is dimensional,
namely the string tension.
The dimensionless gauge and Yukawa
couplings are then set by ratios of the dilaton and moduli field  vacuum
expectation values to the string tension.  Similarly the gravitational
coupling (Planck mass) is scaled from the string tension by a modulus
vev. Thus until these vevs are fixed, the fundamental coupling
constants could vary in time.  Of course, it is widely expected that
non-perturbative effects will generate a potential for the moduli and fix
their vev's (probable at some high energy scale), the mechanism and scale
of this fixing are a subject of much debate. Thus in principle, one can
consider variations in the fundamental couplings a logical possibility.

Indeed, a considerable amount of interest in the possibility of
time-varying constants has been generated by recent observations of
quasar absorption systems. Observations of the energy level splitting
between the
$S_{1/2} \to P_{3/2}$ and $S_{1/2} \to P_{1/2}$ transitions in several
atomic states such as CIV, MgII, and SiIV, suggest a time variation in the
fine structure constant by an amount ${\Delta \alpha / \alpha} = (0.72
\pm 0.18) \times 10^{-5}$ \cite{Webb01} over a redshift range of 0.5 --
3.5. In addition, there may be preliminary evidence for a variation
${\Delta \mu / \mu} = (5.7\pm 3.8)\times 10^{-5}$ in the
ratio of the proton to electron masses $\mu \equiv m_p/m_e$ \cite{ivan}
for redshifts of $\sim$ 2 -- 3.

Starting from the work of Bekenstein \cite{Bek}, there have been a number
of attempts to formulate a
dynamical model of a variable fine structure constant \cite{bar,opv}.
These models typically consist of a massless scalar field which has a
linear coupling to the
$F^2$ term of the $U(1)$ gauge field.
The coupling of non-relativistic matter to the scalar field
induces a cosmological change in the background value of this
field which can be interpreted as a change in the effective fine
structure   constant. Independent of our prejudices
(or lack thereof) regarding a fundamental theory, such models are difficult to
construct in such a way as to remain consistent with other experimental
constraints.  For example,
 the presence of a massless scalar field in the theory
leads to the existence of an additional attractive force which does not
respect Einstein's weak universality principle. The extremely accurate
checks of the latter \cite{EDB} lead to a
firm bound that
confines possible changes of $\alpha$ to the range
$\Delta \alpha / \alpha < 10^{-10}-10^{-9}$
for $0< z <5$ \cite{Bek,Livio,opv} in the context of the minimal
Bekenstein model where a change in the scalar field is triggered by the
baryon energy density. It was argued \cite{opv} that a
significant $O(1)$ coupling  between the scalar field and the dark
matter energy density is required in order to allow
${\Delta \alpha / \alpha} \sim 10^{-5}$ and remain consistent with
equivalence principle constraints \cite{EDB}. Thus it is natural to
expect that in generalized Bekenstein models, not only the fine
structure constant but all of the couplings and masses  will
depend on the expectation value of a light scalar.

In addition, there exist various sensitive experimental checks that
coupling constants do not change (See e.g., \cite{Sister}).
Among the most stringent of these is the bound on
 $|\Delta \alpha/\alpha|$ extracted from the analysis
of isotopic abundances from the Oklo phenomenon \cite{Oklo}-\cite{LV}, a
natural nuclear fission reactor that occurred about 1.8 billion years
ago. While the Oklo bound $|\Delta \alpha/\alpha|< 10^{-7}$ is
considerably tighter than the ``observed" variation, Oklo occurred at a
time period corresponding to a redshift of about 0.14, and it is quite
possible that while $\alpha$ varied at higher redshifts, it has not
varied recently.  That is, there is no reason for the variation to be
constant in time. Big bang nucleosynthesis also provides limits on
$\Delta \alpha/\alpha$ \cite{bbn,co}. Although these limits are weaker,
they are valid over significantly longer timescales.

The Bekenstein model and its modifications are introduced in {\em ad hoc}
manner, and their relation to deeper motivated theoretical models is
problematic. A major stumbling point on the path between the theory and
phenomenology of a changing $\alpha$ is the masslessness of the modulus
that mediates this change
\cite{bd}.
Indeed, to be relevant for the cosmological evolution now
or in the recent past,
the mass of this scalar has to be comparable or
lighter than the Hubble parameter
at $z \sim 0 - 5$, whereas quantum corrections would tend to generate a
much {\em larger} mass. This is a generic problem for any interacting
quintessence-like model that is similar to the cosmological constant
problem.  Since very little is actually understood about the latter, we
do not think  that this problem is a sufficient reason to discard
phenomenological  models of changing $\alpha$. Disregarding the problem of
masslessness of  the modulus that renormalizes coupling constants, we
proceed to  analyze phenomenological constraints on a theory with a fixed
(modulus-independent) high-energy scale $M_P$, unified values for all
coupling constants at $M_P$,  and a single modulus that changes the
values of {\em all} coupling constants.  Such a theory is motivated by a
string model with dilaton-dependent  coupling constants. One has to keep
in mind, however, that the simplest  string tree-level values for the
couplings of dilaton to matter and $m_{dilaton} =0$ lead to a
catastrophic non-universality in the gravitational  exchange by this
scalar, which violates the current bound by 10 orders of magnitude
\cite{opv,DP}. One remedy to this problem may be a more complicated form
of the dilaton-matter coupling with a universal extremum \cite{DP}.
Another  possibility is that a massless modulus contains a relatively
small admixture of the  string dilaton, so that all the couplings of this
modulus to matter are  suppressed to a level consistent with the
equivalent principle \cite{EDB}.

The possibility that significantly stronger constraints on the variation
of the fine structure constant can be obtained in the context of
theories in which the change in a scalar field
vev induces a change in the fine structure constant as well as the
other gauge and Yukawa couplings was first explored in \cite{co} (see
also,
\cite{ds}).  There it was recognized that in any unified theory in which
the gauge fields have a common origin, variations in the fine structure
constant will be accompanied by similar variations in the other gauge
couplings.  In other words, variations of the gauge coupling at the
unified scale will induce variations in all of the gauge couplings at the
low energy scale.
Note that even in theories with
non-universality at the string scale, there is almost always some
relation between the couplings.

It is easy to see that the running of the strong coupling constant has
dramatic consequences for the low energy hadronic parameters, including the
masses of nucleons \cite{co}.
Indeed the masses are determined by the QCD scale, $\Lambda$, which is
related to the ultraviolet scale, $M_{UV}$, by dimensional transmutation:
\be
\alpha_s(M_{UV}^2) \equiv {g_s^2(M_{UV}^2) \over 4 \pi} =
{4 \pi \over b_3\ln (M_{UV}^2/\Lambda^2)},
\ee
where $b_3$ is a usual renormalization group coefficient that depends on
the number
of massless degrees of freedom, running in the loop.
Clearly,  changes in $g_s$ will induce (exponentially) large
changes in $\Lambda$:
\be
{\Delta \Lambda \over \Lambda} = {2 \pi \over 9 \alpha_s(M_{UV})}
{\Delta \alpha_s(M_{UV})\over \alpha_s(M_{UV})} \gg
{\Delta \alpha_s(M_{UV})\over \alpha_s(M_{UV})},
\label{deltaLambda}
\ee
where for illustrative purposes we took the beta function of QCD with
three
fermions. On the other hand, the electromagnetic coupling $\alpha$ never
experiences
significant running from $M_{UV}$ to $\Lambda$ and thus
$\Delta \Lambda / \Lambda\gg\Delta \alpha/\alpha $. A more elaborate
treatment
of the renormalization group equations
%, with varying number of fermions and
%MSSM beta function 
above $M_Z$ \cite{lss} leads to the result that is in
perfect agreement with \cite{co}:
\be
{\Delta \Lambda \over \Lambda}\simeq 30 {\Delta \alpha \over \alpha}.
\label{enhance}
\ee

In addition, we expect that not only the gauge couplings will vary,
but all Yukawa couplings are expected to vary as well.  In \cite{co}, the
string motivated dependence was found to be
\be
{\Delta h \over h} = {\Delta \alpha_U \over \alpha_U}
\ee
where $\alpha_U $ is the gauge coupling at the unification scale and $h$
is the Yukawa coupling at the same scale. However in theories in which
the electroweak scale is derived by dimensional transmutation, changes in
the Yukawa couplings (particularly the top Yukawa) leads to exponentially
large changes in the Higgs vev.
In such theories, the Higgs expectation value
corresponds to the renormalization point and
is given qualitatively by
\be
v\sim M_P \exp (- 2 \pi c / \alpha_t)
\label{rad1}
\ee
where $c$ is a constant of order 1, and $\alpha_t = h_t^2/4\pi$.
Thus small changes in $h_t$ will induce large changes in $v$.
For $c \sim h_t \sim 1$,
\be
{\Delta v \over v} \sim 80 {\Delta \alpha_U \over \alpha_U}
\label{enhance2}
\ee
This dependence gets translated into
a variation in all low energy particle masses.  In short, once we
allow $\alpha$ to vary, virtually all masses and couplings are expected
to vary as well, typically much more strongly than the variation induced
by the Coulomb interaction alone.
 Unfortunately, it is very hard to make a
quantitative prediction for $\Delta v/v$ simply because we do not know
exactly how the dimensional transmutation happens in the Higgs sector,
and the answer will  depend, for example, on such things as the dilaton
dependence of the  supersymmetry breaking parameters. This uncertainty is
characterized in Eq. (\ref{rad1}) by the parameter $c$.
For the purpose of the present discussion
it is reasonable to assume that $\Delta v/v$ is comparable but not
exactly equal to $\Delta \Lambda/\Lambda$. That is, although they are both
$O(10-100) \Delta \alpha/\alpha $, their difference $|\Delta
\Lambda/\Lambda - \Delta v/v|$ is of the same order of magnitude which we
will take as 
$ \sim 50 \Delta \alpha/\alpha$.

In \cite{co}, these relations were exploited to derive a strong bound on
variations of $\alpha$ during big bang nucleosynthesis. The standard
limit \cite{bbn} of ${\Delta \alpha / \alpha} \la
10^{-2}$  is improved
by about 2 orders of magnitude to ${\Delta \alpha /
\alpha} \la 10^{-4}$ as recently confirmed in a numerical
calculation \cite{bbn2}. Here, we will consider the effect of these
relations on the existing Oklo bounds as well as derive new bounds
relating to the long lived $\alpha$- and $\beta$-decaying isotopes,
 $^{147}$Sm and $^{187}$Re; we will also comment briefly on the
 influence of changing the fundamental couplings on $s$-process yields.

Before proceeding, we note briefly that in the class of theories we are
considering, we would predict that the proton-electron mass
ratio is also affected. For example, we would expect that
\be
{\Delta \mu \over \mu} \sim {\Delta \Lambda \over \Lambda} - {\Delta v
\over v}
\ee
{}From Eqs. (\ref{enhance}) and (\ref{enhance2}), we estimate that
${\Delta \mu / \mu} \sim 3 \times 10^{-4}$, based on the reported
claim of a variation in $\alpha$ \cite{Webb01} and is somewhat larger than
that reported in \cite{ivan}. For related discussions see \cite{other}.

\section{The Oklo Bound Revisited}

Approximately two billion years ago, a natural fission reactor was
operating in the Oklo uranium mine in Gabon. Shlyakhter \cite{Oklo}
argued that a strong limit on the time variation of $\alpha$ was
possible by examining the isotopic ratios of Sm in the Oklo reactor.
This suggestion was confirmed by Damour and Dyson \cite{DD}, who performed
a detailed analysis of the isotopic ratios and the effect that varying
$\alpha$ would have on the resonant neutron capture cross section of Sm.
Their analysis provided a bound of
\be
\left|{\Delta \alpha \over \alpha}\right| \la 10^{-7}
\ee
The bound was derived primarily by calculating the shift in the
resonance energy,  $E_r \simeq 0.0973$ eV, of the Sm neutron capture cross
section which is induced by a variation in the Coulomb contribution.
While a full analytical understanding of the energy levels of
heavy nuclei is not available, it is nevertheless possible to
obtain an estimate of the size of an energy shift if the fundamental
parameters of the theory are varied. In particular, it is possible to
identify the over-riding scale dependence of the terms which determine
the binding energy of the heavy nuclei. We perform this estimate in the
context of the Fermi gas model.
We will argue that the Oklo data can
provide a sufficiently strong bound on the variation of $m_q$ and
$\Lambda$, or more precisely, on the variation of $m_q/\Lambda$.

We begin by considering the semi-empirical formula for the binding
energy $B(A,Z)$ of a spherical nucleus with mass number $A$ and
atomic number $Z$ \cite{FGM}:
\be
B(A,Z) = C_V A - C_S A^{2/3} - a_C Z^2 A^{-1/3} +
\delta(A,Z) a_{\rm pair} A^{-\epsilon} + C_d Z^2 A^{-1} +
\Delta_{\rm shell}(A,Z),
\label{baz}
\ee
where the first three terms on right-hand side represent the volume,
surface, and Coulomb contributions, respectively, and the last three
terms represent corrections due to pairing, surface diffusiveness, and
shell structure. The corrections due to asymmetry between neutrons and
protons are included in $C_V$ and $C_S$. The coefficients in Eq.
(\ref{baz}) are given by
\ba
C_V &=& a_V \left[1 - \kappa \left({A-2Z\over A}\right)^2\right],\\
C_S &=& a_S \left[1 - \kappa \left({A-2Z\over A}\right)^2\right],\\
a_C &=& {3\over 5}{e^2\over r_0},\\
C_d &=& {\pi^2\over 2}\left({a_0\over r_0}\right)^2{e^2\over r_0}.
\ea
Numerical values for the pertinent quantities are $a_V=15.68$ MeV,
$a_S=18.56$ MeV, $\kappa=1.79$, $a_C=0.717$ MeV, $a_0=0.546$ fm,
$r_0=1.205$ fm, $a_{\rm pair}=11$ MeV, $\epsilon=0.5$, and
\be
\delta(A,Z)=\left\{\begin{array}{rl}
1&\mbox{for even-$A$ even-$Z$},\\
0&\mbox{for odd-$A$},\\
-1&\mbox{for even-$A$ odd-$Z$}.
\end{array}\right.
\ee
The shell structure coefficients are discussed below.

The Coulomb contribution has a simple interpretation as the
electromagnetic energy stored in a uniformly charged sphere of
total charge $Ze$ and radius $r_0A^{-1/3}$. The volume and surface
contributions can be rewritten as:
\be
C_V A - C_S A^{2/3}=-(\langle T\rangle + \langle V\rangle),
\ee
where $T$ and $V$ represent the kinetic and potential energy,
respectively, of the nucleons. Based on the Fermi gas model
and considerations of nucleon-nucleon interaction potential
\cite{FGM},
\ba
\langle T\rangle &=& T_0
\left[1 + {5\over 9}\left({A-2Z\over A}\right)^2\right]
\left(1-{f_S\over A^{1/3}}\right)A, \label{kin} \\
\langle V\rangle &=&
\left[V_V + V_{\rm sym}\left({A-2Z\over A}\right)^2\right]
\left(1-{f_S\over A^{1/3}}\right)A, \label{pot}
\ea
where
\be
T_0={3\over 10}{k_F^2\over M}\propto {1\over Mr_0^2}
\label{to}
\ee
is the zeroth-order contribution to the total kinetic energy
and the terms with the coefficient $f_S$ represent surface correction.
For a Fermi momentum $k_F=1.36$ fm$^{-1}$, $T_0=23.01$ MeV,
and the other quantities can be obtained as
$V_V=-38.69$ MeV, $V_{\rm sym}=15.29$ MeV, and $f_S=1.184$.

Now consider the reaction
\be
^{149}{\rm Sm}+n\to ^{150}{\rm Sm}+\gamma.
\ee
The $Q$-value of the reaction is
\ba
Q&=&B(150,62)-B(149,62)\nonumber\\
&=&-0.987T_0-0.851V_V-0.245V_{\rm sym}
+1.615a_C+8.16\times 10^{-2}a_{\rm pair}-0.172C_d\nonumber\\
&&+\Delta_{\rm shell}(150,62)-\Delta_{\rm shell}(149,62),
\label{Q}
\ea
where the numerical values for all the terms except for the last two
on the right-hand side
are $-22.71$, 32.93, $-3.75$, 1.16, 0.90, $-0.21$ MeV, respectively.
Comparing the $Q$-value calculated above with the experimental
value of 7.99 MeV gives an estimate of
$\Delta_{\rm shell}(150,62)-\Delta_{\rm shell}(149,62)=0.33$~MeV.
Clearly, changes in $T_0$ and $V_V$ produce the largest effects on
the $Q$-value. Thus not only will our limit be strengthened
(relative to the purely electromagnetic bound \cite{DD}) due to the strong
interaction, but also due to the enhanced sensitivity of the binding
energy relative to the Coulomb term alone.

Since we are lacking a complete analytical theory for the origin of the
terms entering into Eq. (\ref{baz}), it will be sufficient to concentrate
on the dominant kinetic and potential terms.
Even in this simplified formalism, it is not possible to account for the
exact scaling of the dimensionful terms $T_0$, $V_V$ and $V_{\rm sym}$.
However, we can identify a certain degree of required scaling, and
it is quite clear that an exact cancellation of such a scaling is
extremely unlikely. Furthermore,
the resonant cross-section proceeds through an excited state of
$^{150}$Sm, which happens to lie very close to the Q value given in
Eq. (\ref{Q}). Thus the quantity of interest is,
\be
E_r = Q - E^* = 0.0973 {\rm eV}
\ee
where $E^*$ is the energy of the excited state of $^{150}$Sm.
Unfortunately, we expect that variations in the fundamental couplings
also lead to changes in $E^*$. The previous bound on $\alpha$ was based
on the presumption that taking into account the variation of
$E^*$ with $\alpha$ only strengthens the bound, so that a
conservative bound on $\Delta \alpha$ can be traced directly to
$\Delta Q$. Here we will have to rely on the probability that it
is also highly unlikely that both $Q$ and $E^*$ depend on all of the
fundamental parameters in exactly the same way.  We return to this point
below.

Before we derive our bound on possible variations of the gauge couplings
it will be useful to first use Eq. (\ref{Q}) to derive the bound on
$\alpha$ along the lines of \cite{DD}. If we ignore all of the
unification arguments given in the previous section, then the only
clearly identifiable piece with the electromagnetic coupling in Eq.
(\ref{Q}) is the Coulomb term.  Damour and Dyson (1996) \cite{DD} derived
 the
bound
\be
-0.12 {\rm eV} <  \Delta (Q - E^*) = \Delta Q < 0.09 {\rm eV}
\label{erbound}
\ee
where
\be
\Delta Q \equiv Q^{\rm Oklo} - Q^{\rm now}
\ee
For our purposes it will be sufficient to consider the limit
$|\Delta Q| < 0.1$ eV.
The Coulomb term in Eq. (\ref{Q}) is simply
\be
1.6 a_C \simeq 1.16 {\rm MeV} \, \left({\alpha \over \alpha_0}\right)
\ee
where $\alpha_0$ is the present value of the fine structure constant.
Thus
\be
\Delta Q = 1.16 {\rm MeV} \,  \left({\Delta \alpha \over \alpha}\right)
\ee
We can therefore immediately derive the limit
\be
\left|{\Delta \alpha \over \alpha}\right| \la 10^{-7}
\ee
in good agreement with Damour and Dyson.

We next attempt to use the same procedure to derive the limit on the gauge
coupling when the unification argument of the previous section is
included. We first note that if we simply associate all dimensionful
quantities as originating from $\Lambda_{QCD}$, no significant limit is
possible.
In such a naive approach, one would argue that the masses of the light
quarks can be  neglected, and all hadronic parameters such as
$m_N$ and the strength of the nucleon-nucleon
potential scale linearly with $\Lambda$. It is clear, though, that in this
limit there is no sizable effect on the position of the resonance, since
\be
{\Delta E_r} \sim \Delta (Q - E^*) \sim {(Q - E^*) \over \Lambda} \Delta
\Lambda
\ee
and hence
$ \Delta E_r / E_r \sim \Delta \Lambda/\Lambda$. Since the constraint
on variation in $E_r$ is only $O(1)$ (cf. Eq. (\ref{erbound})), one
is simply left with $\Delta \Lambda/\Lambda < O(1)$. This point has been
repeatedly  stressed in the literature. For the most recent discussion,
see, e.g. Ref.
\cite{FS}.

There are two generic problems that prevent a rigorous analysis of
$\Delta E_r$ as  a function of $m_q/\Lambda$.
The first  problem lies in fact on the interface of the perturbative QCD
description  and the description in terms of hadrons. In short, we do not
know the exact dependence  of hadronic masses and coupling constants
on $\Lambda$ and light quark masses. The second problem concerns modeling
nuclear  forces in terms of the hadronic parameters.

Generically, the mass of a hadron $i$ can be parameterized as
\be
m_i \simeq const~\Lambda\left(1+ {\kappa_i} \fr{m_q}{\Lambda} + O(m_q^2)
\right),
\label{m_i}
\ee
where $\kappa_i$ reflects the dependence on the light quark mass.
Clearly, masses of
the members of the lightest pseudoscalar octet have a
significant dependence on
$m_q$, and $\kappa_\pi \sim 1/2$. On the other hand, heavy
hadrons remain massive
in the chiral limit which suggests that for nucleons $\kappa_N \ll 1$.
In practice, $\kappa_N$ (Fig 1a) probably varies from 0.1 to 0.2
due in part to a rather large value of
the matrix element of $m_s \bar ss $ over a nucleon state, $\langle N \vert
m_s \bar ss \vert N\rangle \sim 100 \pm 50 $ MeV~\cite{gs},
which is known to admittedly poor accuracy
(the combined matrix element over $u$ and $d$ quarks is 45 MeV).
Other particles which are important for nuclear dynamics such as
vector  resonances are expected to have $ \kappa_N < \kappa_V <
\kappa_\pi$. We note that although a complete analysis of
$\Delta m_i /\Delta m_q$ is certainly lacking (due to the difficulty of
the problem), some insight can be gained from QCD sum rule analyses.
%%%%%%%%%%%%%%%%%%%%%%%
\begin{figure}
 \centerline{%
   \psfig{file=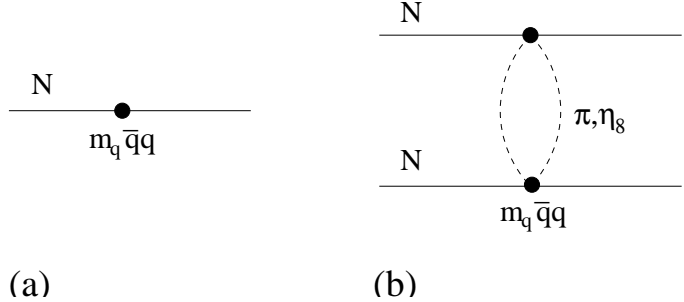,width=9cm,angle=0}%
         }
\vspace{0.1in}
 \caption{The $m_q$ dependence of hadronic parameters. (a) illustrates the
insertion of
 $m_q\bar qq$ into the nucleon line that determines $\kappa_N$. (b) is the
chiral loop that gives the $m_q$ dependence of the nucleon-nucleon
interaction.}
\end{figure}
%%%%%%%%%%%%%%%%%%%%%%%

Recent progress in understanding the chiral dynamics of few nucleon
systems  is not directly applicable to large nuclei such as Sm
(See e.g., \cite{vanKolck} and references therein). Instead, one has to
resort to a  model description of nuclear forces such as contact
interactions generated by  $t$-channel exchanges of vector and scalar
resonances. In this approach,  the relevant Lagrangian can be
symbolically written as
\be
{\cal L} = \sum \bar N (i \partial_\mu \gamma_\mu - m_N) N +
\fr{a}{M^2}(\bar NN)(\bar NN)
 \label{lagr}
\ee
where the sum runs over all relevant isospin and Lorentz structures. The
scale $M$ remains finite in the chiral limit, and thus scales
approximately linearly with $ \Lambda$. It is important to note
that single pion exchange,
which is the most important source of the nuclear force in the deuteron,
plays little  if any role in nuclei with large $A$. Nevertheless, one can
provide a conservative  estimate of the $m_q$ dependence of a
nucleon-nucleon potential by considering chiral  loop corrections to the
contact nucleon-nucleon interaction, Fig. 1b. This way one  obtains
\be
\fr{a}{M^2} = \fr{a_0}{\Lambda^2} +\fr{a_1m_q}{\Lambda^3}+
\fr{1}{8 \pi^2 f_\pi^2}
\left(\fr{\langle N \vert m_q \bar qq \vert N\rangle}{f_\pi}\right )^2 \ln
\left[\fr{4\pi f_\pi}{m_\eta}\right]+ O(m_q^2),
\label{rad}
\ee
and $f_\pi$ scales linearly with $\Lambda$.
In principle, the strength of overall
numerical ($\Lambda$ and $m_q$ independent) coefficient $a$ can be fit to
nuclear data but for the purpose of present discussion, we simply take it
to be $O(1)$. Unfortunately, there is no known reliable way of calculating
the coefficients $a_0$ and $a_1$ in front of the zeroth and first order
terms of the expansion in $m_q$. On the
other hand, it is very unlikely that $a_0$ and $a_1$ would conspire to form
exactly the same combination of $\Lambda$ and $m_q$ as in $m_N$, (\ref{m_i}).
Fortunately, the presence of a chiral
logarithm in (\ref{rad}) in the $O(m_q^2)$ term and the known
{\em absence} of such a term in $m_N$
ensures that the nucleon mass, $m_N$, and the interaction strength,
$1/M^2$, depend differently on
$(m_q/\Lambda)$.
Thus, to be very conservative, we 
consider only the variation of the chiral logarithm in (\ref{lagr}) and
use 100 -- 200 MeV for the value of $\langle N \vert m_q \bar qq \vert
N\rangle$\cite{gs}, noting that the linear term (of order $m_q$) is
likely to produce a more stringent bound:
\be
\Delta {\cal L} \sim (1 - 4) \times 10^{-1} {\rm fm}^{2} (\bar NN)(\bar
NN)
\fr{\Delta(m_q/\Lambda)}{m_q/\Lambda} +...
\label{deltaL}
\ee
{}From $\Delta {\cal L}$ we obtain the Hamiltonian $\Delta H$ using the
mean field  approximation:
\be
\Delta H = (1 - 4) \times 10^{-1} {\rm
fm}^{2}n_0\fr{\Delta(m_q/\Lambda)}{m_q/\Lambda}
\sum_i  \theta(R-r_i),
\ee
where $R$ is a nuclear radius and $n_0$ an average nuclear density
inside the nucleus, $n_0 \simeq 0.17 $ fm$^{-3}$. The sum runs over all
nucleons  inside a given nucleus.
As discussed above, the change in the position of the resonance comes as
the result of the  change in the position of the ground state of
$^{149}$Sm and the excited state of $^{150}$Sm.
%The average of $\Delta H$ in a given nucleus is simply proportional
%to $A$, and our final estimate takes the following form:
In this naive approach, we have neglected all specific origins of the
resonant energy found in $Q$ and $E^*$.  If we then associate $\Delta
E_r$ with the difference in $\Delta H$ between the relevant states
of $^{149}$Sm and $^{150}$Sm, we obtain
\be
\fr{\vert \Delta E_r\vert}{ E_r} \sim \fr{(1 - 4) \times 10^{-1} {\rm
fm}^{2}n_0}{E_r}
\left\vert\fr{\Delta(m_q/\Lambda)}
{m_q/\Lambda}\right\vert \sim (2.5 - 10) \times 10^{7} \left\vert
\fr{\Delta v}{v} -
\fr{\Delta\Lambda}{\Lambda}\right \vert.
\label{deltaer}
\ee

Note, that our result (\ref{deltaer}) is about an order of magnitude
weaker than the result obtained in \cite{FS}.
Applying $\vert \Delta E_r/E_r\vert < 1$
\cite{DD}, we arrive at the following bound:
\be
\left\vert \fr{\Delta v}{v} -
\fr{\Delta\Lambda}{\Lambda}\right \vert < (1 - 4)\times 10^{-8}
\label{lambound}
\ee

In the theoretical framework discussed in this paper, the result
(\ref{lambound}) allows us to improve the limits on the
variation of the coupling constant
by over an order of magnitude compared to Ref. \cite{DD}. Indeed,
combining eqs.  (\ref{lambound}) and the estimate $|\Delta
\Lambda/\Lambda - \Delta v/v| \sim 50 \Delta \alpha/\alpha $, one gets
\be
\left\vert \fr{\Delta \alpha}{\alpha}\right\vert < (2 - 8) \times
10^{-10}
\label{ourbound}
\ee
 We remind the reader that the range quoted in the limit above is due to
the uncertainty in the strange quark contribution to the nucleon mass and
corresponds to using 100 -- 200 MeV for $\langle N\vert m_q \bar qq \vert
N \rangle $.

It is also possible (though it bears its share of uncertainty) to use
the expression for $Q$ (\ref{Q}) to further strengthen the bound.
One should note that the resonance energy, $E_r$, is tiny due to
a cancellation between $Q$ and $E^*$. Both are individually $\approx 8$
MeV. However, the value of $Q$ is also determined by a
set of cancellations between the terms in Eq. (\ref{Q}) which are
of order 20-30 MeV. By once again relying on the improbability that
all terms will depend on the fundamental constant in the
same way, one can in principle use only the largest term (or terms) in $Q$,
which are the potential and kinetic terms.

If we again parameterize our ignorance of exact scalings with
$|\Delta
\Lambda/\Lambda - \Delta v/v|$,
we can expect that
\be
{\Delta E_r \over E_r} \sim {\Delta (Q - E^*) \over E_r} \sim
{\Delta V \over E_r} \sim {V  \over E_r} \kappa \left\vert
\fr{\Delta v}{v} -
\fr{\Delta\Lambda}{\Lambda}\right \vert < 1
\label{strongbound}
\ee
which leads to
\be
\left\vert \fr{\Delta \alpha}{\alpha}\right\vert <(1-5)\times
10^{-10}
\label{ourbound2}
\ee
for the range  $0.1\la\kappa < 0.5 $.
In Eq. (\ref{ourbound2}), we have assumed $V \sim 30$ MeV.  This is the
most optimistic bound that one can expect from the Oklo data.
Recalling, that the Oklo event occurred some 2 Gyr in the past,
we would obtain the limit $\dot \alpha / \alpha < 2.5 \times 10^{-19}$
yr$^{-1}$.

\section{Constraints on the variation of fundamental couplings from
 long lived $\alpha$- and $\beta$-decay nuclei}

Bounds on the variation of the fundamental couplings can also be
obtained from our knowledge of the lifetimes of certain long-lived nuclei.
In particular, it is possible to use relatively precise meteoritic
data to constrain nuclear decay rates back to the time the solar system
was formed (about 4.6 Gyr ago). Thus, for the current standard inputs of
$\Omega_m
\simeq 0.35$,
$\Omega_\Lambda \simeq 0.65$ and $h_0 = 0.71$, we can derive a constraint on
possible variations at a redshift $z \simeq 0.45$ which borders the
range over which such variations are claimed to be observed ($z$ = 0.5 -- 3.5).
Note that nuclei with relatively short lifetimes, especially in
the case of bound beta-decay (see section 4), could be of
astrophysical interest in the context of $s$-process
nucleosynthesis~\cite{ta}.

\subsection{Nuclear physics}

Here, we will concentrate only on the possible influence of variations in
the $Q$-value on $\alpha$- and $\beta$-decay lifetimes. As
fission processes are less well understood and less sensitive than the
other decay processes,  we do not consider those isotopes for which
fission is the dominant decay mode.  On the other hand, $\alpha$- and
$\beta$-decay processes are better understood, and our basic constraints
can be seen as due to a change in the phase space of emitted particles in
the case of $\beta$-decays, while for $\alpha$-decays, the constraints are
related to Coulomb barrier penetrability.

In principle, slight changes in a coupling strength
can stabilize (destabilize) certain isotopes. Clearly,
the maximum effect is expected to occur for nuclei with small $|Q|$.
%($Q< 0$ corresponds to stable isotopes which could
%become unstable if the binding energies and hence $Q$ are slightly altered).
A pioneering study on the effect of variations of fundamental
constants on radioactive decay
lifetimes was performed by Dyson \cite{D}.
The isotopes which are most sensitive to changes in the  $Q$ value
are typically those with the lowest value of $Q_\beta$, which
is the $Q$-value
corresponding to $\beta$-decay. In addition,
${\Delta}Q$'s arising from variations of fundamental
constants are expected to scale with ${\Delta}B$, which is the change in
the binding energy $B$ of the parent nucleus.
Accordingly, as a first step, we consider isotopes with the smallest values of
$|Q_\beta|/B$ as calculated from nuclear mass tables \cite{Audi}, regardless
of their actual decay modes.
For $\alpha$-decay, small values of $Q_\alpha$ (of order the
smallest $Q_\beta$, i.e. a few tens of keV) are not interesting because the
Coulomb barrier is so high that the decay probability is vanishingly small.
Hence, we considered isotopes with the smallest $Q_\alpha$ values which have
dominant $\alpha$-decay mode ($Q_\alpha\sim$ 1~MeV).

The possible influence of the variation of constants on the binding energy
and on the $Q$ values has been discussed in section 2.
Here, we will consider  $\Delta Q/B$ as a parameter
(assuming again that ${\Delta}Q$ scales with ${\Delta}B$, i.e.
${\Delta}Q\propto{\Delta}B$ and
${\Delta}B/B\propto{\Delta\alpha}/\alpha$).
Isotopes with small $Q$ for $\beta^+$-decay are not considered as the
electron capture channel is already open and will dominate the decay.
Tables 1 and 2 present respectively the {\it a priori} most interesting
nuclei concerning $\beta^-$-decay and electron capture (EC). These
isotopes  have $|Q|/B$ less than the typical value of $10^{-4}$ and are
selected from  the NUBASE files of nuclear data \cite{Audi}.
For each isotope, the $Q$ value, lifetime, binding energy $B$
and the $Q/B$ ratio are displayed; also shown is the difference in spin and
parity between the parent and the daughter nuclei, which governs the
degree of forbiddenness. Stable nuclei which could become
unstable are also included.
In Table 3, we show the $\alpha$-decay isotopes with half-life
longer than 10$^6$ years and $1.9<Q_\alpha<4.7$~MeV.

\begin{table}
\caption{Properties of selected nuclei with
 low $|Q_{\beta}/B|$}
%\tiny
\begin{center}
\begin{tabular}{|c|c|c|c|c|c|c|c|c|}
\hline
 A & Element &$Q_{\beta^-}$ &half life &decay&B &$Q/B$&$\Delta I$&$\Delta\pi$\\
   &         &(keV)& (year) && (MeV) &&&\\
\hline
 106&  Ru & 40.0&1.02&$\beta^-$& $9.07\times10^{2}$ &$4.41\times10^{-5}$&1&+\\
\hline
107& Pd&33.0&$ 6.5\times10^{6}$&$\beta^-$&$9.16\times10^{2}$&$3.60\times10^{-5}$&2&-\\
\hline
123&Sb & -53.3 & -.- & stable & $1.04\times10^{3}$
&$-5.11\times10^{-5}$& 3&+\\
\hline
150&Nd & -87.0 &  $2.1\times10^{19}$ & $2\beta^-$&  $1.24\times10^{3}$
&$-7.03\times10^{-5}$& 1&-\\
\hline
151&Sm&76.7 & 90. & $\beta^-$&$ 1.24\times10^{3}$ &$6.16\times10^{-5}$& 0&-\\
\hline
148&Eu & 41.0 &  0.15 & $\beta^+$ &  $1.22\times10^{3}$
&$3.36\times10^{-5}$& 1&-\\
\hline
157&Gd&-60.1&  -.- &stable&$1.29\times10^{3}$&$-4.67\times10^{-5}$&0&-\\
\hline
160&Gd&-105.6& -.-& stable&$1.31\times10^{3}$&$-8.07\times10^{-5}$&3&-\\
\hline
163 &Dy&-3.0& -.-&stable&$1.33\times10^{3}$& $-2.26\times10^{-6}$&1&+\\
\hline
 171 &Tm &96.4&1.92 &$\beta^-$ &$ 1.39\times10^{3}$&$6.96\times10^{-5}$& 0&-\\
\hline
179 &Hf&-110.9& -.-&stable&$1.44\times10^{3}$&$-7.71\times10^{-5}$&1&+\\
\hline
184&Re & 31.5 &  0.10 & $\beta^+$ &  $1.24\times10^{3}$
&$-7.03\times10^{-5}$& 1&-\\
\hline
187 & Re &2.6& $4.35\times10^{10}$&$\beta^-$&$ 1.49\times10^{3}$&$1.74
\times10^{-6}$&2&-\\
\hline
\end{tabular}
\end{center}
\end{table}

\begin{table}
\caption{Properties of selected nuclei with
 low $Q_{EC}/B$}
%\tiny
\begin{center}
\begin{tabular}{|c|c|c|c|c|c|c|c|c|}
\hline
 A & Element &$Q_{E.C.}$ &half life &decay&B &$Q/B$&$\Delta I$&$\Delta\pi$\\
   &         &(keV)& (year) && (MeV) &&&\\
\hline
 107&Ag&-33.0 & -.-&stable&$9.15\times10^{2}$&$-3.60\times10^{-5}$&2&-\\
\hline
123&Te&53.3&$>
6.0\times10^{14}$& EC+ &$1.04\times10^{3}$&
$5.12\times10^{-5}$&  3 & +\\
\hline
136 &Cs&80.0&$3.61\times10^{-2}$& $\beta^-$&$ 1.14\times10^{3}$&$7.01\times10^{-5}$&5&+\\
\hline
150&Pm&87.0&$3.06\times10^{-4}$& $\beta^-$ &$1.24\times10^{3}$
&$7.04\times10^{-5}$ & 1 & - \\
\hline
151 &Eu& -76.7& -.- &stable &$1.24\times10^{3}$& $-6.16\times10^{-5}$&0&-\\
\hline
 157& Tb & 60.1 &71. &  EC  +&$1.29\times10^{3}$& $4.67\times10^{-5}$&0&-\\
\hline
 160 &Tb & 105.6 &0.2& $\beta^-$ &$ 1.31\times10^{3}$&  $8.07\times10^{-5}$&3&-\\
\hline
 163&Ho & 3.0&  4573. & EC  +&$ 1.33\times10^{3}$&$ 2.26\times10^{-6}$&1&+\\
\hline
171&Yb&  -96.4& -.- &stable&$1.38\times10^{3}$&$-6.96\times10^{-5}$
&0&-\\
\hline
 176 &Lu& 106.2& $3.78\times10^{10}$ & $\beta^-$&$1.42\times10^{3}$&$7.49\times10^{-5}$&7&-\\
\hline
 179&Ta& 110.9&  1.82 &  EC  +&$  1.44\times10^{3}$&$  7.71\times10^{-5}$&1&+\\
\hline
178 &W & 90.0&$ 5.92\times10^{-2}$& EC  + &$1.43\times10^{3}$& $6.30\times10^{-5}$&1&+\\
\hline
184&Os& -31.5 & -.- & stable & $1.47\times10^3$& $-2.14\times10^{-5}$&
3 & - \\
\hline
187&Os&  -2.6& -.- & stable&$ 1.49\times10^{3}$&$-1.74\times10^{-6}$&2&-\\
\hline
 193&Pt&  56.6& 50.& EC  +&$ 1.53\times10^{3}$&$ 3.70\times10^{-5}$&1&-\\
\hline
 194 &Hg & 40.0 &440.& EC  +&$ 1.54\times10^{3}$ &$2.61\times10^{-5}$&1&-\\
\hline
202 & Pb &49.0 &$5.25\times10^4$&EC+&$1.59\times10^3$&$3.08\times10^{-5}$&
2 & - \\
\hline
 205&Pb& 51.1& $1.53\times10^{7}$&  EC + &$ 1.61\times10^{3}$&$ 3.17\times10^{-5}$&2&-\\
\hline

\end{tabular}
\end{center}
\end{table}

In the following, we concentrate on the most favorable cases and we study
the variation of their half-life as a function of ${\Delta}Q/B$ in the
limit  of $|{\Delta}Q|/B<10^{-4}$.
We see that the isotope with the smallest $Q_{\beta^-}$ value is
$^{187}$Re ($2.66\pm0.02$~keV).
In the case of electron capture (Table 2) a few isotopes also
have small $|Q|$ values, such as $^{187}$Os and
$^{163}$Ho. These low Q isotopes could be involved in bound
state $\beta$-decay prevailing in stellar conditions (see section 4).

\begin{table}
\caption{Properties of selected nuclei with
 small $Q_\alpha$}
%\tiny
\begin{center}
\begin{tabular}{|c|c|c|c|c|c|c|}
\hline
 Z&A & Element &$Q_{\alpha}$ &half life &B &$Q/B$\\
   &      &   &    (MeV)&      (year) &      (MeV) &\\
\hline
60 &144 &Nd& 1.905&    $2.29\times10^{15}$& $   1.20\times10^3$&
$1.59\times10^{-3}$\\
\hline
 62 &146& Sm& 2.528   &$ 1.03\times10^8$ &$   1.21\times10^3$&  $
2.09\times10^{-3}$\\
\hline
  62 &147& Sm &   2.310& $1.06\times10^{11}$ & $ 1.22\times10^3$&
$1.90\times10^{-3}$\\
\hline
62 &148 &Sm & 1.986& $ 7.\times10^{15}$& $   1.23\times10^3 $&  $
1.62\times10^{-3}$\\
\hline
64 &150& Gd  &  2.809  &$ 1.79\times10^6$&   $ 1.24\times10^3 $&  $
2.27\times10^{-3}$\\
\hline
64& 152& Gd&    2.204 & $  1.08\times10^{14}$& $  1.25\times10^3$&   $
1.76\times10^{-3}$\\
\hline
66& 154 &Dy  &  2.947 &$   3.\times10^6$& $   1.26\times10^3$&   $
2.34\times10^{-3}$\\
\hline
72 &174& Hf&    2.495 &$   2.0\times10^{15}$& $   1.40\times10^3$&  $
1.78\times10^{-3}$\\
\hline
76& 186& Os&    2.822&$  2.0\times10^{15}$& $ 1.48\times10^3 $&$
1.90\times10^{-3}$ \\
\hline
78& 190& Pt&    3.249&$    6.5\times10^{11}$& $1.51\times10^3$&  $
2.15\times10^{-3}$\\
\hline
90&232&Th& 4.082&$1.41\times10^{10}$&$1.77\times10^3$& $2.31\times10^{-3}$\\
\hline
92& 235& U&     4.678&$7.04\times10^8$&$1.78\times10^3$& $2.62\times10^{-3}$\\
\hline
92& 238&U& 4.270&$4.47\times10^9$& $1.80\times10^3$& $2.37\times10^{-3}$\\
\hline
\end{tabular}
\end{center}
\end{table}

\begin{figure}
%\vspace*{-1.8in}
\hspace*{-.30in}
\begin{minipage}{8in}
\epsfig{file=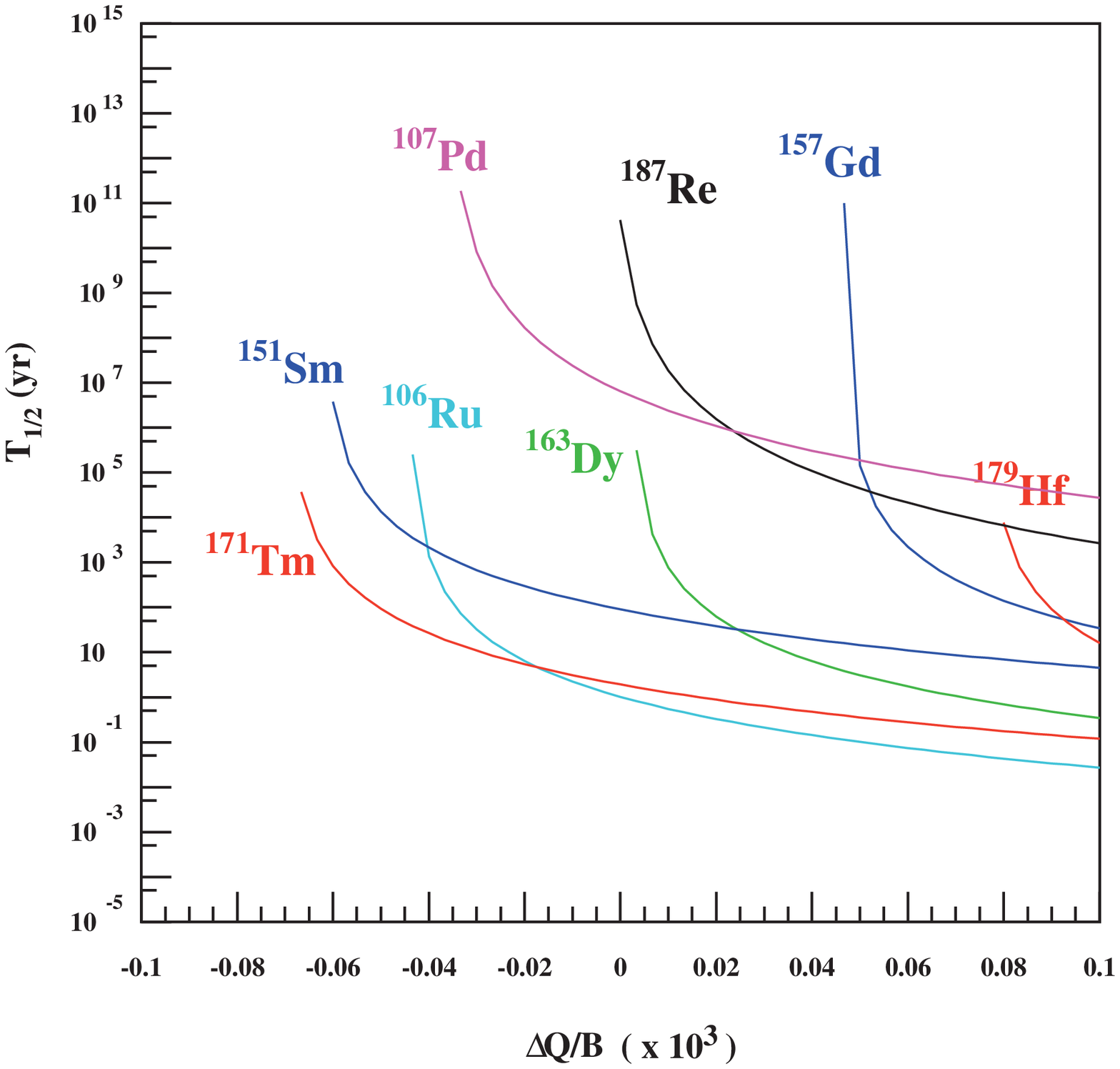,height=3.5in}
\epsfig{file=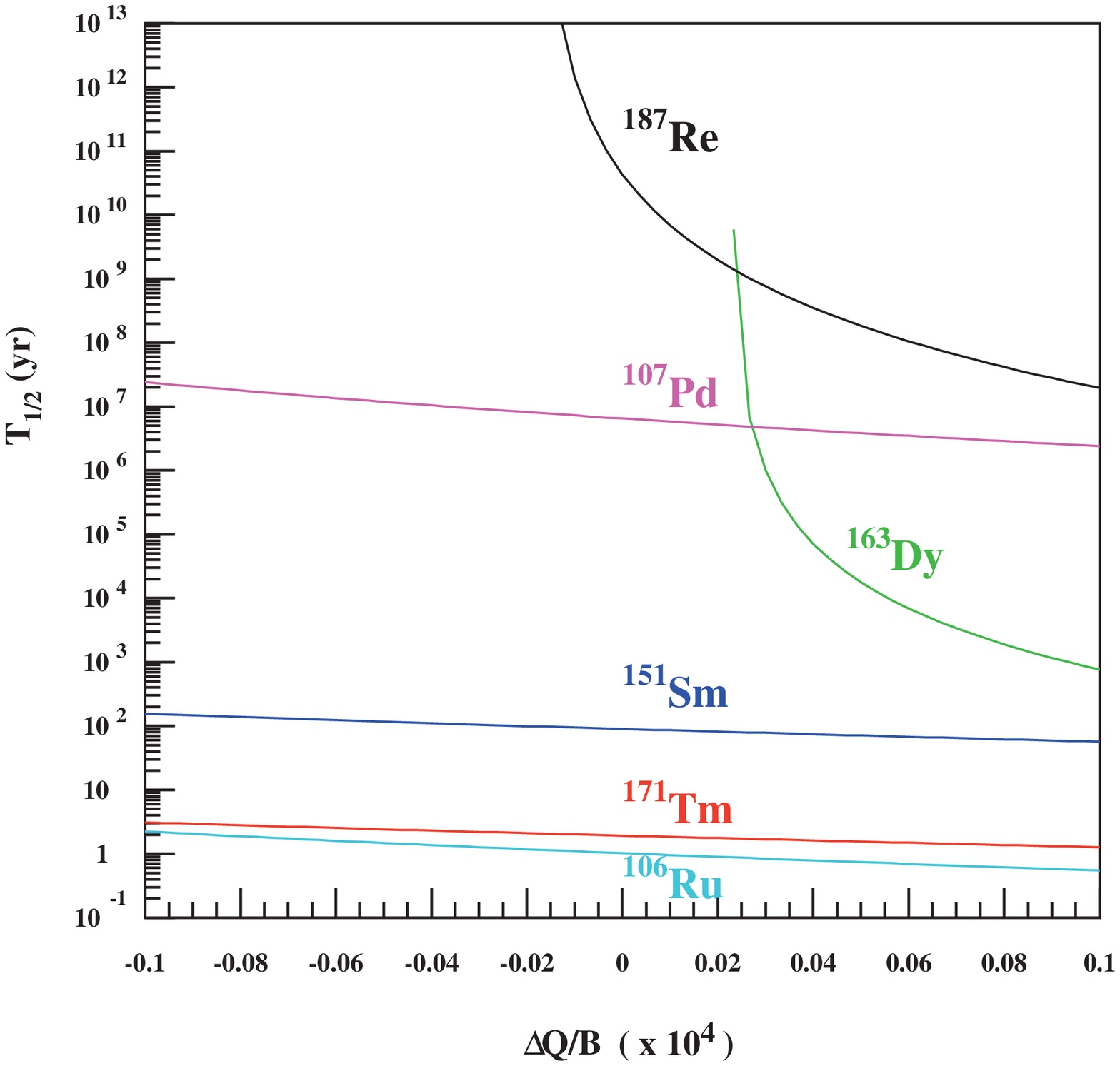,height=3.5in} \hfill
\end{minipage}
\begin{minipage}{8in}
\hspace*{-.30in}
\epsfig{file=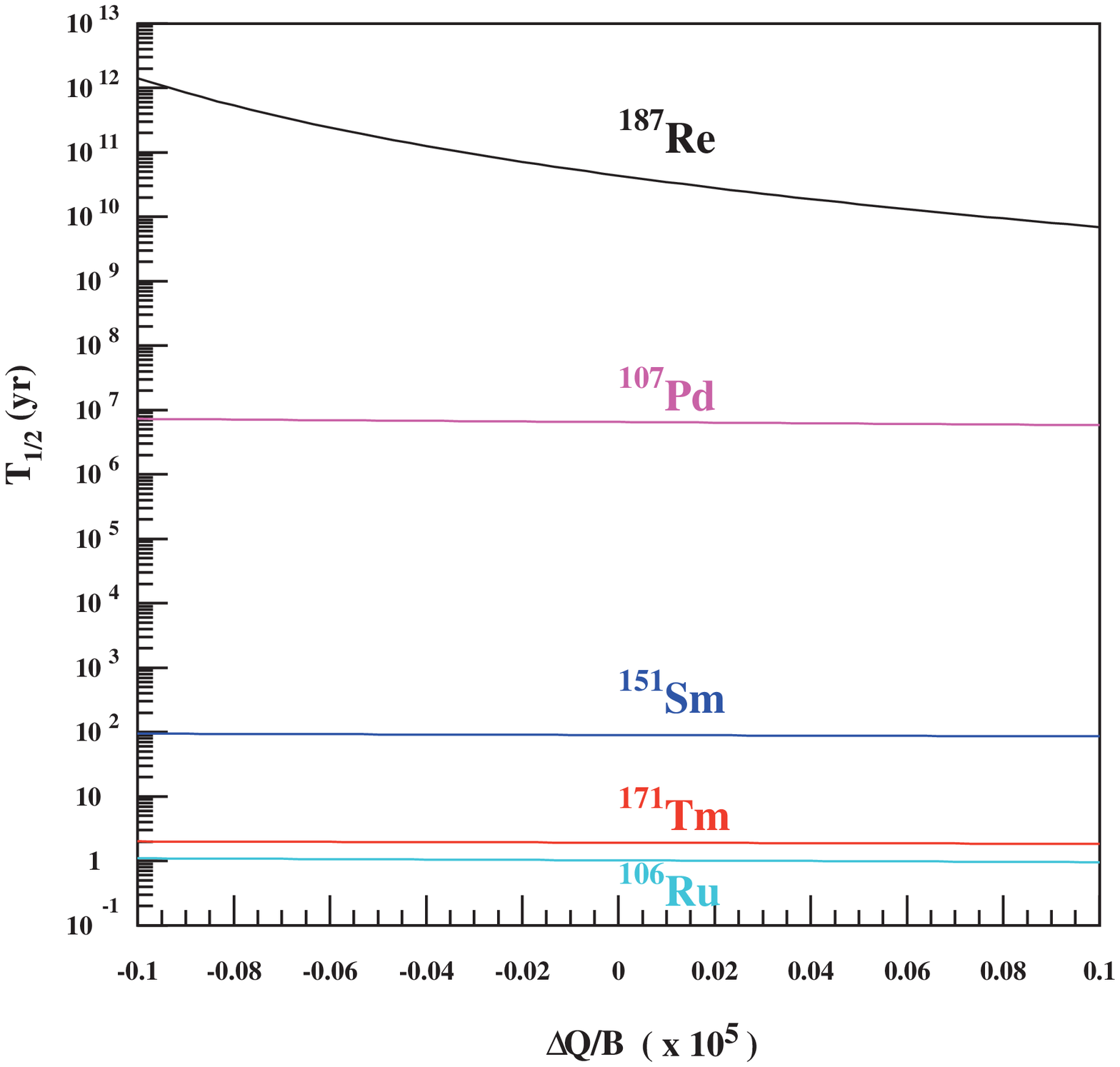,height=3.5in}
%\hspace*{0.5in}
\epsfig{file=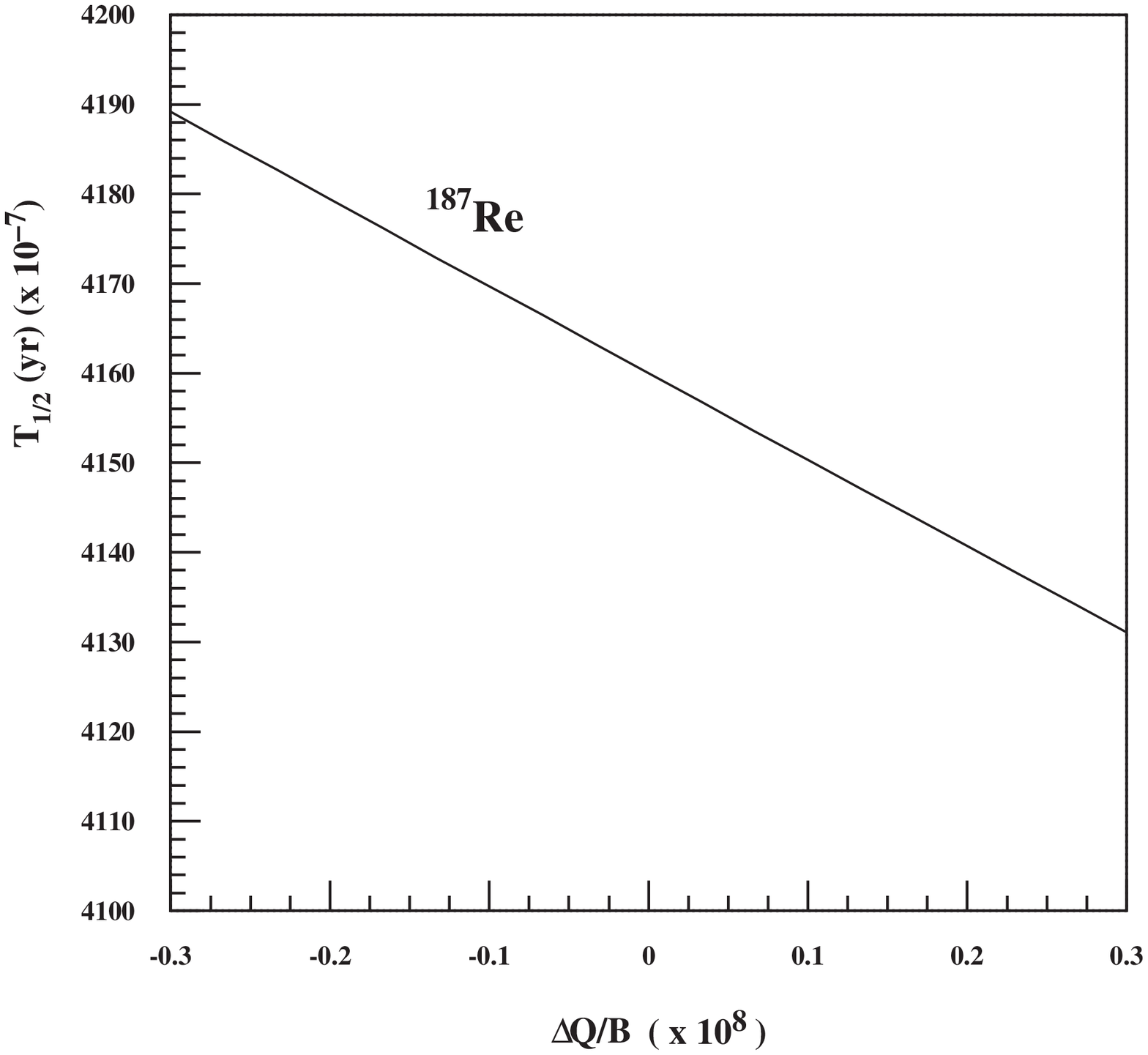,height=3.4in} \hfill
\end{minipage}
%\vskip-0.7in
\caption{\label{fig:Q/B}
{\it
Evolution of half-life of the $\beta^-$-isotopes as a function of the
 $\Delta Q/B$ parameter. In a) the scale is taken to be $\Delta Q/B <
10^{-3}$, in b) $\Delta Q/B <
10^{-4}$, in c) $\Delta Q/B <
10^{-5}$, in d) $\Delta Q/B <
10^{-8}$. }}
\end{figure}

The rate of beta-decay depends on the decay energy and on the leptons'
angular momentum $\ell$. The decays are classified as allowed ($\ell=0$) and
$\ell$--forbidden ($\ell\neq$0). {\em Unique} transitions occur when only
 one multipolarity is permitted by spins and parities.
We will consider here only the allowed and first forbidden transitions and
neglect those that would give an even longer half-life.
The half-live, $T_{1/2}$, is related to the nuclear matrix element,
$M_{fi}$,  through
\be
fT_{1/2}={2\pi^3\ln2\over (m_e)^5G_F^2|M_{fi}|^2},
\ee
where $G_F = 1/2v^2$ is the weak coupling constant and
$f$ contains all the energy dependence.
For allowed transitions, $f$ is given by \cite{DeShalit}
\be
f(Z,R,\epsilon_0)\equiv\int_1^{\epsilon_0}\rho(Z,R,\epsilon_e)
\left(\epsilon_0-\epsilon_e\right)^2\epsilon_e\sqrt{\epsilon_e^2-1}
\;\;{\rm d}\epsilon_e
\ee
where $Z$ is the atomic number of the daughter nucleus,
$\epsilon$ ($\epsilon_0$) is the (maximum) electron
energy in units of $m_e$, and
\be
\rho(Z,R,\epsilon_e)=2(2p_eR)^{2(s-1)}(1+s)e^{\pi\eta}
\left|{{\Gamma(s+i\eta)}\over{\Gamma(2s+1)}}\right|^2
\ee
is the relative electron density at the nuclear
surface (radius $R$), with $\eta\equiv{\alpha}ZE_e/p_e$ and
$s^2\equiv1-(Z\alpha)^2$. We will use $f$ to calculate the energy
dependence of the half-life for the allowed and
first forbidden transitions. For unique first forbidden
transitions ($\Delta I=2$, $\Delta\pi=-$), we will use the
approximation $f_u\approx(\epsilon_0^2-1)f$ (see \cite{SF}).
The calculated half-life is displayed in Figure \ref{fig:Q/B}
as a function of ${\Delta}Q/B$ for the isotopes of interest.
The individual panels of Figure \ref{fig:Q/B} correspond to
$|{\Delta}Q|/B<10^{-n}$ for $n= 3$, 4, 5, and 8.

When $|\Delta Q|/B$ is relatively large as in Fig.
\ref{fig:Q/B}a, we see that several isotopes may change from stable to
unstable or vice versa. For smaller variations as seen in
Fig. \ref{fig:Q/B}b,
we see that the half-life is slightly altered for a few isotopes,
but the most spectacular
effects arise for $^{187}$Re and $^{163}$Dy, the latter becoming unstable.
When we restrict variations to $|\Delta Q|/B< 10^{-5}$,
we see that only the $^{187}$Re half-life is significantly altered.
After an examination of Tables 1, 2, and 3 and the dependence
of the half-life on $\Delta Q$, we conclude that
$^{187}$Re is the most promising
isotope for studying variations of fundamental constants.
We will discuss $^{187}$Re in more detail after considering the case of
long-lived $\alpha$-decays.

\subsection{Limits due to Long lived $\alpha$-decays}
Dyson \cite{D}
defined the sensitivity of the decay constant of the nucleus
to the change of the electromagnetic coupling constant as
\be
s = {\alpha \over \lambda} {d\lambda \over d\alpha},
\ee
which is a function of the decay energy $Q$.
The $\alpha$-decay rate depends
on $\alpha$ through the probability of Coulomb barrier penetration.
Approximating this rate by
\be
\lambda \propto \exp\left[-{4\pi\alpha Z\over\sqrt{Q_\alpha/(2m_p)}}\right],
\ee
where $Z$ is the atomic number of the daughter nucleus, we can write
\be
s = -{4\pi\alpha Z\over\sqrt{Q_\alpha/(2m_p)}}\left(1 - {1 \over 2}u\right),
\ee
where
\be
u = {\alpha \over Q_\alpha} {dQ_\alpha \over d \alpha}.
\ee
The decay energy $Q_\alpha$ is given by
\be
Q_\alpha=B(A,Z)+B_\alpha-B(A+4,Z+2),
\ee
where $B_\alpha$ is the nuclear binding energy of the $\alpha$-particle.
Varying only the Coulomb terms in Eq. (\ref{baz}) and neglecting
the contributions from $B_\alpha$, we obtain $s=774$, 890, 575, 659,
571, 466, and 549 for $^{147}$Sm, $^{152}$Gd, $^{154}$Dy,
$^{190}$Pt, $^{232}$Th, $^{235}$U, and $^{238}$U, respectively.

The variation $\Delta\alpha/\alpha$ is related to
$\Delta\lambda/\lambda$ as
\be
{\Delta\alpha\over\alpha}={1\over s}{\Delta\lambda\over\lambda}.
\ee
Since the values of $s$ for the nuclei listed above are similar,
the most stringent constraint on $\Delta\alpha/\alpha$ is given by
the nucleus with the smallest known $\Delta\lambda/\lambda$.
While it is tempting to use $^{238}$U for this purpose, 
the uncertainty of $6.7\times 10^{-4}$ given in Table 4, corresponds to
a laboratory measurement and is not directly applicable to 
a constraint at high redshift. Indeed, $^{238}$U is extremely well 
measured, and is used to calibrate the ages of the meteorites.
Instead, we consider $^{147}$Sm as an example, and assume that
$\Delta\lambda/\lambda$ is less than the fractional meteoritic
uncertainty of $7.5\times 10^{-3}$ in the half-life of $^{147}$Sm
\cite{be1} given in Table 4 (see also \cite{be}). This gives
\be
{\Delta\alpha\over\alpha} \la 10^{-5}.
\ee

\begin{table}
\caption{Sensitive and insensitive $\alpha$- and $\beta^{-}$-isotopes}
%\tiny
\begin{center}
\begin{tabular}{|c|c|c|c|c|c|}
\hline
Nucleus& Decay& $Q$ (MeV)& Half-life (yr)& $\Delta\lambda/\lambda_{\rm
lab}$ (\%) & $\Delta\lambda/\lambda_{\rm
met}$ (\%) \\
\hline
$^{238}$U&$\alpha$&4.27&$4.468\times10^9$&$0.067$ & --\\
\hline
$^{235}$U&$\alpha$&4.678&$7.038\times10^8$&$0.071$ & --\\
\hline
$^{232}$Th&$\alpha$&4.082&$1.405\times10^{10}$&$0.4$ & $> 1$ \\
\hline
$^{147}$Sm&$\alpha$&2.31&$1.06\times10^{11}$&$ 1.9$ & 0.75\\
\hline
$^{87}$Rb&$\beta^{-}$&0.283&$4.75\times10^{10}$&$0.8$  & $0.8$\\
\hline
$^{40}$K&$\beta^{-}$&1.311&$1.265\times10^9$&$0.6$  & $1$\\
\hline
\end{tabular}
\end{center}
\end{table}

The change in $\lambda$ due to more general variations of the fundamental
constants can be written as
\be
{\Delta\lambda\over\lambda}=-{4\pi\alpha Z\over\sqrt{Q_\alpha/(2m_p)}}
\left[{\Delta\alpha\over\alpha}+{1\over 2}\left({\Delta m_p\over m_p}-
{\Delta Q_\alpha\over Q_\alpha}\right)\right].
\label{dlamb}
\ee
We again take $^{147}$Sm as an example, for which
\ba
Q_\alpha&=&3.36T_0+3.40V_V-0.070V_{\rm sym}+39.9a_C
-0.975C_d\nonumber\\
&&+\Delta_{\rm sh}(143,60)
-\Delta_{\rm sh}(147,62)+B_\alpha,
\ea
where the numerical values for all the terms except for the last three
on the right-hand side are 77.31, $-131.51$, $-1.08$, 28.63,  $-1.18$
MeV, respectively. The experimental value of $B_\alpha$ is 28.30 MeV.
Comparing the $Q_\alpha$ calculated above with the experimental value of
2.31 MeV gives an estimate of $\Delta_{\rm sh}(143,60)
-\Delta_{\rm sh}(147,62)=-1.84$ MeV. We again see that changes in $T_0$
and $V_V$ produce the largest effects on $Q_\alpha$.
Note however, that because the decay-rate depends on the ratio of $Q$ to
$m_p$, if we neglect quark mass contributions to $Q$ and $m_p$ and retain
only the scaling due to $\Lambda$, we see that the contributions from
$\Delta m_p / m_p$ and $\Delta Q / Q$ in Eq. (\ref{dlamb}) cancel and no
improvement in the limit is possible. Including the quark contribution
through the coefficient, $\kappa$, we expect that the  scaling of
$T_0$ and
$V_V$ with
$m_q/\Lambda$ will not be exactly the same as that of $m_p$. So Eq.
(\ref{dlamb}) gives
\be
{\Delta\lambda\over\lambda}\simeq
-{2\pi\alpha
Z\kappa\over\sqrt{Q_\alpha/(2m_p)}}{T(V)
\over Q_\alpha} {\Delta(m_q/\Lambda)\over(m_q/\Lambda)}
=(3 - 20)  \times 10^2\left\vert
\fr{\Delta v}{v} - \fr{\Delta\Lambda}{\Lambda}\right \vert,
\label{alphabound}
\ee
using ${\Delta(m_q/\Lambda)/(m_q/\Lambda)} = \left\vert
{\Delta v}/{v} -
{\Delta\Lambda}/{\Lambda}\right \vert$.
The range is due to the difference in $T$ and $V$ (80 -- 130 MeV)
compared to $Q \sim 2$ MeV and the range in $\kappa = 0.1 - 0.5$ as
discussed above. The variation in Eq. (\ref{alphabound}) corresponds to
\be
{\Delta\alpha\over\alpha}<(0.8 - 5) \times 10^{-7}
\ee
for $\Delta\lambda/\lambda<7.5\times 10^{-3}$.
As shown below,  this constraint is less
stringent than that derived from $^{187}$Re $\beta$-decay.

\subsection{Limits due to Long lived $\beta$-decays}

In section 3.1, we have seen  that $^{187}$Re is the most sensitive
indicator of a possible variation of $\alpha$ as first argued by Peebles
and Dicke \cite{PD} and Dyson  \cite{D}.
The upper limit obtained from the analysis of the Re/Os ratio in
iron meteorites  obtained at that time was  $\dot
\alpha/\alpha  = 5\times10^{-15}$ yr$^{-1}$
\cite{D}, and was less stringent than the Oklo limit \cite{DD}. In spite
of this, Re is of interest since the estimate of the effect of the
variation of the coupling constants (particularly when we go beyond
variations in $\alpha$) based on the resonance energy (in the Oklo
case) is more complicated  than that based on the Q value (in the Re
case). As we saw in section 2, the constraints based on the Sm resonant
energy required some knowledge of the role of quark masses in the nucleus
and a limit based on variations of $\Lambda_{QCD}$ alone could not be
obtained. In the case of $\beta$-decay,
a constraint can be obtained in a more direct
way.   Above all, the Re analysis is independent of the
Oklo analysis and uses
different physics. Finally, the dramatic improvement in the meteoritic
analyses of the $^{187}$Re/$^{187}$Os ratio mandates an update of the
constraint on variations of the coupling constants.

Rhenium occurs in relatively high concentration in iron rich meteorites.
The $^{187}$Re decay rate has been determined through the generation
of high precision isochrons from material of known ages, particularly
iron meteorites. Using the Re-Os ratios of IIIAB iron meteorites that
are thought to have been formed in the early crystallization of asteroidal
cores, Smoliar et al. \cite{smo}  (see also \cite{fas}) found a
$^{187}$Re half-life of 41.6 Gyr within 0.5\% assuming that the age
of the IIIA iron meteorites is 4.5578 Gyr $\pm 0.4$ Myr which is identical
to the Pb-Pb age of angrite meteorites \cite{lug}. For a general
discussion see  refs. \cite{bi,be}. The results of Smoliar et al.
\cite{smo} are in good agreement with those of Shen et al.  \cite{shen},
which adds confidence to the meteoritic value of the half-life (which is
more precise than the direct measurement
\cite{lin} which carries a 3\% uncertainty). The ages of iron meteorites
determined by Rhenium dating are in excellent agreement with other
chronometers such as U-Pb and Mn-Cr, which means that the Rhenium
lifetime has not varied more than 0.5\% over the age of iron meteorites
(4.56 Gyr). This gives
$\dot\lambda/\lambda< 1.1\times 10^{-12}$ yr$^{-1}$, to be
compared with the limit of $10^{-10}$ yr$^{-1}$ given by Dyson \cite{D}.
The improvement in the data is considerable.

The $\beta$-decay of $^{187}$Re is a unique first fordidden transition,
for which the energy dependence of the decay rate can be approximated as
\cite{RE}
\be
\lambda\propto G_F^2 Q_\beta^3 m_e^2.
\ee
which is in agreement with \cite{D} and gives a good description of the
variation of $T_{1/2}$ with $Q_\beta$ shown in Fig. \ref{fig:Q/B}d. The
decay energy,
$Q_\beta$, is given by
\ba
Q_\beta&=&B(187,76)-B(187,75)+(m_n-m_p-m_e)\nonumber\\
&=&0.339T_0+0.611V_{\rm sym}-26.4a_C+0.807C_d+\Delta_{\rm sh}(187,76)
-\Delta_{\rm sh}(187,75)\nonumber\\
&&+(m_n-m_p-m_e),
\ea
where $m_n$ is the neutron mass and the numerical values for all 
terms
except for the last three
on the right-hand side are 7.80, 9.34, $-18.93$, and 0.98 MeV,
respectively.
The experimental value of $(m_n-m_p-m_e)$ is 0.78 MeV.
Comparing the $Q_\beta$ calculated above with the experimental value of
2.66 keV gives an estimate of $\Delta_{\rm sh}(187,76)
-\Delta_{\rm sh}(187,75)=0.03$ MeV. Considering only the variation of
the
Coulomb term in $Q_\beta$, we have
\be
{\Delta\lambda\over\lambda}=3{\Delta Q_\beta\over Q_\beta}\simeq
3\left({20\ {\rm MeV}\over Q_\beta}\right)
\left({\Delta\alpha\over\alpha}\right)\simeq 2\times 10^4
\left({\Delta\alpha\over\alpha}\right),
\ee
which gives $\delta\alpha/\alpha<3\times 10^{-7}$ for
$\delta\lambda/\lambda<0.5\%$ over a period of 4.6 Gyr or
$\dot\alpha/\alpha < 6 \times 10^{-17}$ yr$^{-1}$.
This is 100 times more stringent than the constraint in \cite{D}
due to the improvement of the limit on $\dot\lambda/\lambda$.

The contributions to
$Q_\beta$ from $T$ and $V$, which scale with $\Lambda$, are comparable to
that from the dominant Coulomb term, which scales as $\alpha \Lambda$.
As changes in $\Lambda$ are $O(30)$ times larger than that in $\alpha$,
we can estimate
\be
{\Delta\lambda\over\lambda}=3{\Delta Q_\beta\over Q_\beta} -  2
{\Delta v \over v} \simeq 3{T(V,C)\over
Q_\beta}{\Delta\Lambda\over\Lambda}\simeq
2 \times 10^4{\Delta\Lambda\over\Lambda},
\label{Rhlimit}
\ee
which gives
\be
{\Delta\alpha\over\alpha}< 8 \times 10^{-9}
\ee
for $\Delta\lambda/\lambda<0.5\%$ over a period of 4.6 Gyr or
$\dot\alpha/\alpha < 2 \times 10^{-18}$ yr$^{-1}$.

Note that all of these
limits based on Re decay hold only if the variation of $\lambda$ is of
the same order as the accuracy of $\lambda$. This hypothesis can be
cross-checked by different chronometric pairs with different
sensitivities to variations of $\alpha$.
Furthermore, we can check a posteori, that even though the meteoritic 
ages are determined in part by lab measurements of the $^{238}$U
lifetime, the limits above still hold. To see this, we note that
the adopted uncertainty in the Re half-life is determined by the 
uncertainty in the slope of $^{187}$Os/$^{188}$Os vs.
$^{187}$Re/$^{188}$Os. The uncertainty in 
the age of the meteorites is neglected. However, Re is far more 
sensitive to changes in $\alpha$ than is U (ie., the sensitivity factor
for U is about 500, while for Re it is 2 $\times 10^4$). 
It is relatively simple to check that a consistent limit requires changes
in $\alpha$ which are sufficiently small so that the uncertainty in
the meteoritic age can be neglected. 

\section{ $s$-process nucleosynthesis}

A variation of the fundamental constants could have several other
significant consequences on astrophysical processes particularly
on nucleosynthesis. For example, variations in the gauge couplings would
affect the position of the triple $\alpha$-resonance necessary for the
synthesis of
$^{12}$C. This has been examined recently by Oberhummer et al. \cite{ob}.
Here, we focus on the nucleosynthesis of heavier isotopes and in
particular those generated by the $s$-process. We will keep our discussion
qualitative since the nucleosynthesis of neutron rich isotopes is complex
at both nuclear and stellar levels.
Branching on the $s$-process path occurs every time the $\beta$-decay
lifetime of a given isotope is commensurate with the neutron capture
lifetime (see
 \cite{ka},
 for a review). Among the nuclei listed in table 1,  several species are
involved in $s$-process branching (specially  in the $Sm-
 Eu-Gd$, $W-Re-Os$ and $Hg-Tl-Pb$ regions, on the basis of their
low $Q/B$  ratio \cite{be2,kaa,ar,yoyo}). $s$-process nucleosynthesis has
been studied in the context of the classical
 constant temperature scenario and more realistically in connection to
 thermal pulses in AGB stars. In the constant temperature scenario of the
 classical $s$-process, thermal excitation of low lying energy states
takes  place, strongly
 modifying stellar lifetimes, and the subtle effects of the variation
of
 the coupling constants are masked. However, more realistic models of
the stellar
$s$-process invoke highly convective situations during thermal pulses
followed
 by long episodes of quietness. The AGB recurrent thermal pulses give rise
 to rapid mixing of freshly synthesized material in cooler zones. In these
 periods, the effect of variations in $\alpha$ could show up, since in
the interpulse regime
 the thermal population of excited levels is suppressed.  Thus, the
situation is involved and deserves a dedicated analyses on the basis of
refined nuclear networks coupled
 to realistic stellar models.

There is however an interesting case which benefits from the high
 temperatures involved, that of metastable $^{180}$Ta, which is the
rarest isotope in nature and  is assumed to be predominantly of
$s$-process origin
 \cite{yo,wi}, (see however \cite{ra}). It is also the only isotope which
is stable in the isomeric state.
 Would it remain metastable if the coupling constants were different in the
past?
 This question also deserves investigation. Here, we will assume that it
does remain metastable.
 The difficulty in producing this isotope in
contrast to the facility of its destruction by thermally
 induced depopulation of the short lived $^{180}$Ta ground state in stars
is  reflected in its
rarity.
 The population of an excited state of $^{179}$Hf at relatively high
 excitation energy ($7/2-$ at 214 keV) is crucial to the synthesis of the
 metastable $^{180}$Ta.
The decay rate of excited $^{179}$Hf, in turn, is essentially dominated
at high temperature
 by the bound state $\beta$-decay, thus a small alteration of its Q value
(say by a few keV)
 would have strong consequences on the final Ta yield (at least in the
 classical $s$-process context).
As the $Q_\beta$ value for decay from the 214 keV excited state of
$^{179}$Hf is $\approx 100$ keV, the variation of $\Delta
Q_\beta/Q_\beta$
would be limited to less than a few percent. Comparing this with
$\Delta Q_\beta/Q_\beta=(1/3)\Delta\lambda/\lambda<2\times 10^{-3}$
for $^{187}$Re, we expect that the constraints
derived from considerations of $^{179}$Hf decay would be $\sim $ 10 times
weaker than those presented in section~3.3 for $^{187}$Re.

More generally, bound state  $\beta$-decay is expected to occur
in highly ionized media in which the
decay electron has a high probability to be captured in an empty atomic orbit.
It is the time reversed process of orbital (bound) electron capture. It occurs
whenever
 Q ($ <0$) is of the order of the binding energy of electrons in the innermost
shells.
$^{187}$Os with $Q = -2.6 keV$ is particularly sensitive to variations
of the fundamental couplings. Other
interesting cases are $^{121}$Te, $^{163}$Dy, and $^{205}$Tl.
 Thus, in stellar conditions, new
disintegration channels could open up and stable nuclei in the laboratory
 could
become unstable. Consequently, a slight perturbation of Q values could have
significant consequences on the results of the $s$-process and more
importantly on the Re/Os datation  \cite{argo}. Indeed, the $^{187}$Re
decay rate could be considerably enhanced in stellar interiors \cite{yo}
by the bound beta-decay of highly ionized $^{187}$Re. At typical
$s$-process temperatures ($3. 10^8$ K), the bound state $\beta$-decay of
$^{187}$Re in to the 9.75 keV
 $^{187}$Os level
is energetically possible, provided the degree of ionization is high. The
 non
unique forbidden transition may give an overwhelming contribution to the
$^{187}$Re decay rate, non unique transitions being, in general much
faster than unique ones. This effect could be easily suppressed by a
slight change of
 $Q$ related
to a change of couplings.

\section{Summary}

We have considered the class of unified theories in which gauge and Yukawa
couplings are determined dynamically by the vacuum expectation value
of a dilaton or modulus field. While such theories may allow the
possibility that the fundamental coupling constants are variable (in
time), they general do so in an interdependent way.  That is, one expects
on quite general grounds that a variation in the fine structure constant
is accompanied by a variation in all gauge and Yukawa couplings.
Even more importantly, as described in \cite{co}, such variations are
also accompanied by variations in quantities such as $\Lambda_{QCD}$ and
the Higgs expectation value which are determined from the gauge and
Yukawa couplings by transdimensional mutation. Limits on the variations in
these dimensionful quantities impose severe bounds on variations in
the fine-structure constant. In the context of dynamical models where the
change in fundamenntal parameters is goverened by a nearly massless
modulus, these limits are complementary to the constraints imposed by
checks of the equivalence principle.

Within this context, we have re-examined the constraints which can be
obtained from the natural nuclear reactor at Oklo.
The previous bound \cite{DD} of $\Delta \alpha / \alpha < 10^{-7}$ was
found by limiting the variations in the Coulombic contribution of
the $Q$ value for the resonant neutron capture process. We
found that by including variations in $Q$ which are induced by
variations in
$\Lambda$ and the quark masses, $m_q$, we can improve this
bound by approximately two to three orders of magnitude.
The improvement is due to 1) the sensitity of both $\Delta \Lambda /
\Lambda$ and $\Delta v / v$ to $\Delta \alpha / \alpha$ (a factor of 50),
and 2) the sensitivity to the dominant terms kinetic and potential terms
in $Q$ rather than the Coulomb term (a factor of 30). However, we lose a
factor of 2 -- 5, due to the rather uncertain contributions of the
quark masses to the nuclear potential in a heavy nucleus.
Thus we obtain the limit  $\Delta \alpha / \alpha < (1 - 5)
\times 10^{-10}$. It is clear that this bound, which is valid at the time
period corresponding to a redshift $z \simeq 0.15$, would require severe
fine-tuning in any model which attempts to fit the recent quasar
absorption data with $\Delta \alpha / \alpha \sim 10^{-5}$ over redshifts
$z = 0.5 - 3.5$. In particular, our results strengthen the need for
the fine-tuning in Bekenstein-type models, as emphasized in \cite{opv}.

We have also considered the bounds which can be derived from long-lived
or barely stable isotopes. Small variations in $\alpha$ ($\Lambda$ and
$v$) can lead to changes in the $Q$ value in heavy nuclei.
We showed that limits from $\alpha$-decay nuclei such as $^{147}$Sm can be
as strong as  $\Delta \alpha / \alpha <  10^{-5}$ from purely Coulombic
variations, and  $\Delta \alpha / \alpha < (0.8 - 5) \times 10^{-7}$
based on $^{147}$Sm life-time uncertainties as small as $\Delta
\lambda / \lambda < 7.5 \times 10^{-3}$.

We further showed that improvements in meteoritic abundance
determinations have enabled one to derive substantially stronger bounds
based on the $\beta$-decay lifetimes of $^{187}$Re. From the analysis of
the Re/Os ratio in meteorites with ages of 4.56 Gyr (known to an accuracy
of 0.01\%), the half-life of $^{187}$Re has been determined to an
accuarcy of about 0.5\%. Purely Coulombic variations in $\alpha$ lead to
the bound $\Delta \alpha / \alpha < 3 \times 10^{-7}$.  From the age of
the meteorites, this limit is applicable at redshift $z = 0.45$.
Thus not only is it competitive with the Oklo bound numerically, but
it corresponds to a higher redshift further emphasizing the difficulty in
achieving a value of $\Delta \alpha / \alpha \sim 10^{-5}$ at higher
redshift.
We further stress that the physics involved in deriving this bound is
independent to that used for the Oklo bound.  When variations in
$\Lambda$ are included, the bound is improved to $\Delta \alpha / \alpha
< 8 \times 10^{-9}$.

\noindent{ {\bf Acknowledgments} } \\
\noindent
  We warmly thank J. L. Birck, G. Manhes and M.
Arnould for providing us with information about meteoritic
 analyses, and P. Vogel, M. Galeazzi, and W. B\"uhring for communications
 on the energy dependence of $^{187}$Re decay. We also thank M. Srednicki
for useful conversations. This work was supported in part by DOE grants
DE-FG02-94ER-40823, DE-FG02-87ER40328, and DE-FG02-00ER41149
at the University of Minnesota and by PICS 1076 CNRS
France/USA. The work of M.P. is supported by P.P.A.R.C.

%%%%%%%%

\end{document}